\documentclass[conference]{IEEEtran}
\ifCLASSINFOpdf
   \usepackage[pdftex]{graphicx}
   \DeclareGraphicsExtensions{.pdf,.jpeg,.png}
\else
\fi

\usepackage{cite}
\usepackage[final]{changes}
\definechangesauthor[name={Dominik}, color=red]{dominik}
\definechangesauthor[name={Suat}, color=blue]{suat}
\definechangesauthor[name={Mumin}, color=purple]{mumin}

\usepackage{atbegshi,picture}
\usepackage{lipsum}

\AtBeginShipout{\AtBeginShipoutUpperLeft{%
  \put(\dimexpr\paperwidth-1cm\relax,-1.5cm){\makebox[0pt][r]{\framebox{This paper was presented in IEEE MASS REU workshop, 2019, San Diego, CA}}}%
}}

\begin{document}

%
\title{Assuring the Integrity of Videos from Wireless-based IoT Devices using Blockchain}

\author{\IEEEauthorblockN{Dominik Danko\IEEEauthorrefmark{1},
Suat Mercan\IEEEauthorrefmark{2}, Mumin Cebe\IEEEauthorrefmark{2}, and Kemal Akkaya\IEEEauthorrefmark{2}} 
\IEEEauthorblockA{\IEEEauthorrefmark{1}Dept. of Math and Computer Science, Clark University, Worcester, MA 01610\\ Email: ddanko@clarku.edu}
\IEEEauthorblockA{\IEEEauthorrefmark{2}Dept. of Elec. and Comp. Engineering, Florida International University, Miami, FL 33174\\ Email: \{smercan,mcebe,kakkaya\}@fiu.edu}

}

\maketitle

\begin{abstract}
Blockchain technology has drawn attention from various communities. The underlying consensus mechanism in Blockchain enables a myriad of applications for the integrity assurance of stored data. In this paper, we utilize Blockchain technology to verify the authenticity of a video captured by a streaming IoT device for forensic investigation purposes. The proposed approach computes the hash of video frames before they leave the IoT device and are transferred to a remote base station. To guarantee the transmission, we ensure that this hash is sent through a TCP-based connection. The hash is then stored on multiple nodes on a permissioned blockchain platform. In case the video is modified, the discrepancy will be detected by investigating the previously stored hash on the blockchain and comparing it with the hash of the existing frame in question. In this work, we present the prototype as proof-of-concept with experiment results. The system has been tested on a Raspberry Pi with different quality of videos to evaluate performance. The results show that the concept can be implemented with moderate video resolutions. 

\end{abstract}
\begin{IEEEkeywords}
Video Integrity; Blockchain; IoT device; hyperledger; digital forensics
\end{IEEEkeywords}

%
\IEEEpeerreviewmaketitle

\section{Introduction}
Video scene is a very important material to interrogate a crime and resolve any dispute because it reveals so much detailed information about the case \cite{kim, poisel}. The proliferation of IoT devices enables convenient video recording opportunities which can be quickly transferred through the availability of various wireless communication options. This may include drones which are used in many smart city applications as well as other wireless-based cameras deployed on streets/buildings, including those used by police officers in pursuing crimes \cite{Edwin}\cite{Hamid}.  

However, video forgery techniques are so sophisticated that a video is susceptible to much manipulation, such as being falsified with insertion or deletion of objects. Advanced video editing tools allow users to tamper \added[id=dominik]{with} videos easily and \added[id=dominik]{in} visually undetectable ways\cite{sanjary}. In particular, if the video is being transmitted through wireless channels, this may become a more prevelant issue as untrusted communication channels might also be the source for data tampering \cite{liang1}.

\begin{figure*}[h]
\begin{centering}
\fbox{
\includegraphics[trim= 50 220 50 20,clip,scale=0.6]{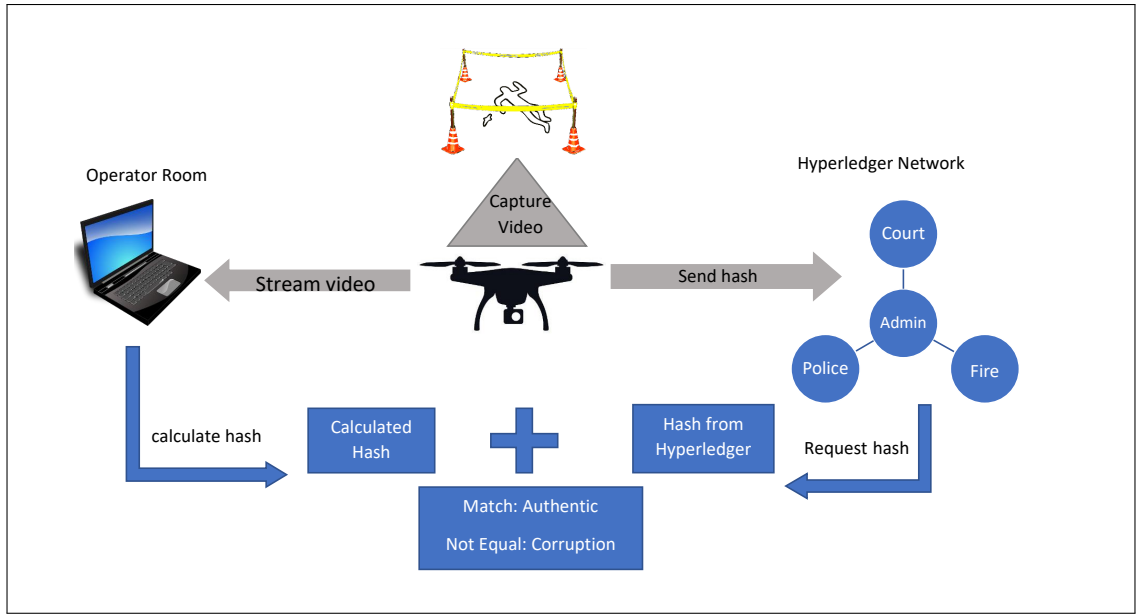}}
\caption{System Design}
\vspace{-6mm}
\label{fig:diagram}
\end{centering}
\end{figure*}

Differentiating a tampered video from an authentic one is now harder than tampering with it. In order to use a video as evidence in court, the authenticity must be proven.  Therefore, it is important to equip the camera with integrity verification abilities. There exist various methods to achieve this goal in the literature. One way is using a digital watermarking technique,\cite{watermarking} in which an invisible signature is inserted into the video that can be used to check if it has been modified. Another approach is analyzing the content itself to detect distortions in the image \cite{wang}. Hashing has also been in use for long time, where a hash is calculated and distributed along with the video. While these approaches may address the issue, there are risks when the data is stored on a server. Data stored on servers may be subject to unauthorized alterations, especially if the process of digital evidence gathering and storage is not followed properly. In addition, if streaming is ongoing, the video frames are transmitted separately, which may require individual attention.  Therefore, there is a need to ensure the integrity of video at all levels, from capturing to storage, so that it can be permissible as evidence in courts, even if the authenticity of the video is questioned as being fake or tampered.  

Blockchain has emerged as an alternative method to transfer money between non-trusting participants without having a trusted third party such as a bank \cite{bitcoin}. The idea of creating a distributed public ledger has led to many other applications and blockchain has exceeded its original purpose\replaced[id=dominik]{;}{ as} some people consider it a game changer. The main idea is to store any data in a distributed and retroactively unchangeable manner which will ensure the authenticity of data. 

In this paper, we apply \deleted[id=dominik]{the} this promising idea to verify the video recorded by wireless-based IoT devices, which can be used in many scenarios. For example, police officers need to carry a wearable camera on their shoulder or forehead which records incidences, and the video footage they capture can contain evidence that would be useful in court.\deleted[id=dominik]{, especially when they are on operations.} 
Similarly, drone videos of crime or accident scenes can contain forensic evidence. Drones are also capable of transmitting real-time video to a base station. 

However, videos can be tampered with during transmission or while stored on a server or other media. Thus, it is necessary to verify if a video has been changed. This fact has motivated us to apply blockchain to address the issue. Specifically, the video data gathered from an IoT device needs to be stored on Blockchain immediately as it leaves the IoT device. This will mitigate risk of tampering by eliminating intermediate stages. Since Blockchain’s consensus mechanism distributes the data to all stakeholders, everyone will have the original copy, and thus no one can modify the data by themselves.

Nevertheless, there are challenges in this approach. Storing the whole video on \added[id=dominik]{a public} blockchain is not free due to transaction fees and blocksize limits. Thus, it is necessary to calculate the hash of each video frame and only write this hash to save space. We propose using a reliable communication protocol to ensure that the hash values will not be lost and thus advocate the use of TCP protocol while the original video data can be sent using an unreliable protocol such as UDP. 

This is still not a viable option as there are thousands of frames in videos. Therefore, we opt for a private (i.e., permissioned) Blockchain approach where the distributed ledger is maintained by a group of stakeholders who are permitted to become members of the private group. In such a case, the proof of work for transaction verification will be different, and the costs will be eliminated. Therefore, we propose using IBM's Hyperledger for our purposes that will act as a distributed storage for our frame hash values.

We implemented the proposed mechanism on a Raspberry Pi that has a camera using WiFi connection to a server. The results indicate that the mechanism is feasible and does not interfere with the performance of the real-time streaming. 

This paper is organized as follows: In the next section, we summarize the related work in the literature. Section III provides some background on the used concepts while Section IV presents the system model along with our approach. In Section V we assess the performance of the proposed mechanism. Section VI concludes the paper.

\section{Related Work} 
Some recent works in the literature started to investigate the validation of the integrity of video data for different purposes using blockchain. 
The main work in this context is presented in \cite{blocksee} which tries to protect not only video content but also camera settings such as angle of camera in surveillance systems. The authors try to prevent hackers from changing camera orientation which might either violate the privacy of neighbors or prevent the recording of some criminal scenes. They distinguish the background and foreground images. Background is used to deduce the camera settings by using some features that do not change over time such as corners and edges, while the foreground is used to identify events occurring in the scene. The hash of the video and metadata is then stored on blockchain. This work's main concern is the parameter settings and more importantly it assumes the availability of the whole video to get the hash. Our goal in this work is different as we deal with video streaming where integrity depends on the reliability of each frame as it leaves the IoT device. Hashing needs to be done in real-time, which puts burden on the IoT device. 

The other closely related work is reported in \cite{liang} where the authors try to provide data assurance for the collected data through IoT sensors. They calculate the hash of the data and store it on the blockchain network instead of the whole data. This work differs from ours as theirs focuses on light text data instead of video data, and they do not deal with wireless communication. 


\section{Background}
\label{sec:background}

\underline{Blockchain}: Blockchain is a list of records called blocks, first proposed by Satoshi for Bitcoin \cite{bitcoin}. These blocks are linked together by containing the hash of the previous block, while containing the data of the current block. The list of blocks continues to grow with the addition of new ones as it is not possible to delete existing blocks. A critical component of blockchain is that all of these blocks, and the data they contain, are distributed among many different nodes. These nodes have to agree on the state of the blockchain, making it nearly impossible to modify any data that has been written to a blockchain. \added[id=mumin] {This working scheme of blockchain carries unique properties such as relieving central authority trust, immutability, and timestamping.
}
These powerful properties are why blockchain is useful and appropriate to use in verifying the authenticity of data and video in our case.

\underline{Consensus Mechanism}:  \replaced[id=mumin]{ The process of adding a new block to the chain is carried out via a protocol, which establishes consensus among participants to confirm the new block.
In other words, it validates the transactions within the block and provides an agreement on the last state of blockchain. There are two types of blockchain structure, public and permissioned, according to the used consensus mechanism\cite{okada2017proposed}.
The most widely-known blockchains, such as
Bitcoin 
and Ethereum 
fall into public blockchain category where consensus is established via a mechanism called Proof-of-Work (PoW). The PoW consensus is typically  a form of hash puzzle which requires finding a predefined hash value.
This consensus protocol brings a significant level of security on the chain (withstand up to 50\% of nodes are being malicious), but at the cost of computational power and time. For instance, Bitcoin's maximum throughput is  7 transactions per second and the consensus finality can take an hour. On the other hand, permissioned blockchains utilize some kind of Byzantine fault tolerant voting based algorithm as consensus mechanism, such as Practical Byzantine Fault Tolerance (PBFT) 
\cite{castro2002practical} 
or Stellar Consensus Protocol (SCP)
\cite{mazieres2015stellar}, which do not require computationally expensive hash puzzles. 
As a result, reaching a consensus is faster which means higher transaction throughput. However, permissioned blockchains generally require more than two-thirds of nodes to be trustworthy rather than 51\%.}{The main idea is to create a mechanism that can prove the legitimacy of data without a having central authority \cite{consensus}. There are many different types of consensus mechanisms that are popular for use in blockchain networks. Some of these include Proof of Work, Proof of Stake, Proof of Identity, and Proof of Elapsed Time. Consensus mechanism is important for the blockchain because it determines the way that the peers are communicating and deciding which transactions are legitimate. The most popular is Proof of Work, which is adapted by Bitcoin and others. In Proof of Work, there are miner nodes that have to perform complex computations to approve transactions. While this works well, it is very resource heavy and expensive.}

\deleted[id=mumin]{Proof of Stake is an alternative to Proof of Work where it doesn't consume so much resources for mining blocks and performing complex calculations, so it is environment friendly. The peer who will create the next block is selected randomly based on some criteria such as wealth or age. Proof of Elapsed Time assigns random waiting times to nodes. The node whose wait time finishes first produces the next block.}

\begin{figure}
    \centering
    \includegraphics[width=\linewidth]{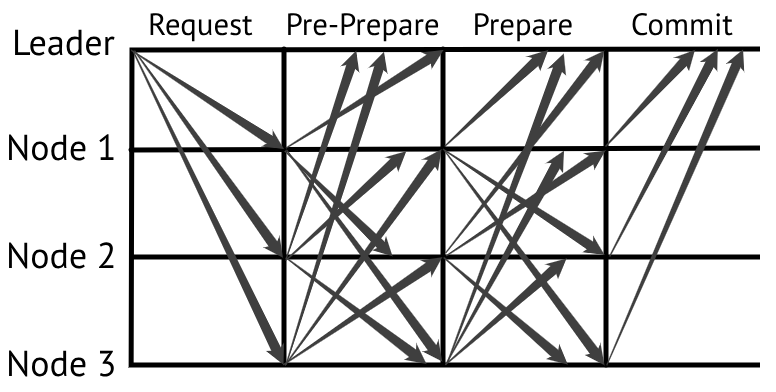}
    \caption{An illustration of how BA protocol works with replicated nodes.}
    \label{fig:PBFT}
\end{figure}

\underline{Voting based consensus protocol}: Voting based consensus protocol first emerged in distributed computing \cite{castro2002practical} to  provide reliability of data or computation, even if arbitrary nodes conduct malicious actions or fail. Permissioned blockchain mechanisms adapt the same idea  to establish consensus where some of some nodes may act maliciously. In this setting, where there are $n$ nodes, a consensus can be achieved if at least $(2n-1)/3$ number of nodes act honestly. Honesty means providing correct information to the other participants. In a permissioned blockchain, there are two different types of nodes called \textit{Leader} and  \textit{Validator}. First, a randomly selected Leader builds a block from transactions. This block is then distributed by the leader to Valitador nodes for verification. Validators check the transactions within the block, sign it, and distribute it again to the other Validators, as shown in Fig.~\ref{fig:PBFT}. Each node, again, distributes the block captured from the other node. This continues until each Validator node collects individually signed versions of the block from the other ones. After $n-1$ version of blocks are gathered, the Validators check differences between blocks. If $(2n-1)/3$ of these blocks are valid, the Validator nodes inform the Leader about confirmation and add the block to its local chain.

\underline{Hyperledger}: Hyperledger\cite{hyperledger} is an opensource platform, distrubuted ledger founded by Linux Foundation and supported by over 50 companies including IBM, Intel. It is implementing permissioned blockchain technology by utilizing a voting based consensus protocol called  Practical Byzantine Fault Tolerance (PBFT). It is a permissioned blockchain platform where access is restricted to stakeholders unlike the public blockchain where \replaced[id=mumin]{anyone can access the produced blocks.}{the data is distributed to all participants. Using a public blockchain like Bitcoin or Ethereum would cost a lot of money because of the high mining cost required by their consensus mechanisms.} A \replaced[id=mumin]{permissioned}{private} blockchain would make more sense for this use case, where only certain entities can access and modify the data stored on the blockchain. \deleted[id=mumin] {has to be considered with caution, however. If only certain entities have access to the blockchain, it makes the data much easier to manipulate, defeating the purpose of storing data on the blockchain so it is immutable. The solution to this is to use a private blockchain that has enough participants. Only certain organizations will be allowed to participate in the network.
}

\section{Proposed Approach}
\label{sec:pre}

We propose a method that leverages the blockchain concept to improve video integrity and to detect tampered video captured by drones. The idea can be accomplished by storing the video frames on a blockchain, which would make it immutable: no one can change it, thus it can be used as evidence for forensic purposes. However, video frames are too big to put on a blockchain, so we take hashes of video frames and put them on a blockchain. These hashes are immutable and can later be used to verify the video. They are sent from the IoT device directly to the blockchain to prevent any alteration during data transmission. 
Fig. \ref{fig:diagram} illustrates the approach that we propose in detail using a drone video communication application. The drone equipped with a camera captures a video frame by frame.

\subsection{Hash Computation}
The hash calculations occur on the IoT device itself so that any intervention by adversaries with the video during transmission will be detected. The hash is calculated either for each frame or for a sequence of frames before it is transferred. However, hashing each frame and sending it to blockchain might be costly in terms of computation and time. Thus, we try various optimization methods to reduce the total time and to handle higher quality videos. 

To get the hash of a frame, the frame is first converted to a string value and then a hash is calculated on this string. Therefore, conversion from frame to string and hashing the string can take a lot of time, especially if this is done for each and every frame. As this will limit the quality of the videos that can be processed, we propose to perform frame selection by adjusting the number of frames we select for hashing. In this respect, one potential approach is to focus on I-frames which are the frames that cannot be encoded using popular encoding techniques such as MPEG or H.264. As the number of such frames would be lower, this will reduce the computation time and eliminate the necessity to store each frames' hash. When a video is encoded, it is converted to GOPs (Group of Pictures) which consists of I, B and P frames. Only I-frames are complete images, B and P frames reflect the changes from surrounding I-frames.

\subsection{Communication of the Hashes}
Since the hash of a frame is a crucial element to be stored in blockchain, it needs to be transmitted in a reliable manner. Therefore, we propose that the IoT device will open a TCP connection to hyperledger so that any loss value could be re-transmitted through the wireless channel. The video frames are typically sent via UDP as it is better suited for video streaming, although this may cause some of the frames to be lost. Therefore, each frame ID will also be appended to the hash in order to be able to compare it with the hash computed for the frame at the server. Any missing IDs will be discarded on the blockchain.  

\subsection{Writing to Blockchain}
In a permissioned Blockchain, there need to be members. For this work, we assumed that there may be three peer organizations in the hyperledger network: 1) Related court unit; 2) Police Department;  and 3) Local fire departments. Note that the number of participants can be increased depending on the nature of the application if needed, but to enable any proof of work algorithm, it should not be less than three. These members will execute the Hyperledger transaction verification process through voting. If there are fewer than three participants, then it would not make sense to use a permissioned blockchain, a public blockchain should be used instead; however, the transaction costs associated with a public blockchain must be considered.
 

\subsection{Integrity Verification}

Later when it comes to use this video as evidence in court, the same hashing process is repeated for the stored video frames. Using the frame number as an index, each frame can be queried in the hyperledger which will return the original hash computed and stored in hyperledger. If the hashes match, the frame is authentic; if they do not match, it can be inferred that the video is a fake or altered video. The stored hash is secure because it is distributed on all stakeholders and they agree on its correctness. If any stakeholder is compromised, the other nodes will still provide the correct information.

\section{Performance Evaluation}

\subsection{Experiment Setup}
\begin{figure}[h]
\begin{centering}
\includegraphics[scale=0.6]{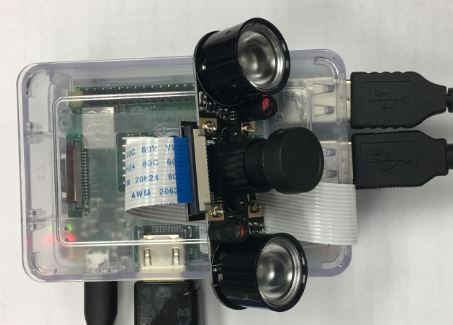}
\caption{System Implemenentation}
\label{fig:imp}
\end{centering}
\end{figure}

In order to evaluate the performance of the proposed approach, we set up a testbed and performed various tests. \replaced[id=dominik]{We used a Raspberry Pi3 to simulate a drone, or similar IoT device. It ran the code that would be installed on a drone equipped with a camera. This setup is illustrated in Fig~\ref{fig:imp}. To ensure consistency in our tests, we stored prerecorded videos of various resolutions on the Raspberry. We also set up a hyperledger network with three participants. The hyperledger network runs on a laptop and communication is achieved through Wi-Fi.}{We have used a Raspberry Pi3 to run the code which would be installed on a drone equipped with a camera, prerecorded videos in various resolutions, and a hyperledger network with three participants. The hyperledger runs on a laptop and communication is achieved through Wi-Fi.} The performance greatly depends on the hardware. Raspberry Pi3 B+ specs shown in Table \ref{table2} are used to run the experiments. Therefore, the values in the results are specific to this configuration. We used OpenCV \cite{opencv} to process the frames. And we employed MD5 hash function for hashing purposes. 

\begin{table}[h]
\renewcommand{\arraystretch}{1.5}
\caption{Raspberry Pi 3 specifications}
\label{table2}
\centering 
\begin{tabular}{|c|c|}
\hline
 CPU(SoC) & BCM2837B0 quad-core A53 1.4GHz  \\
\hline
\hline
 RAM & 1GB LPDDR2 SDRAM  \\
\hline
\hline
 Networking & 2.4GHz and 5GHz 802.11b/g/n/ac Wi-Fi  \\
\hline
\end{tabular}
\end{table}

We used 6 different resolutions of the same prerecorded video to be precise in measurements. It is a 10 second video in mp4 format which has 303 frames in total. MPEG is used as the encoding function. The size and resolution of each version is shown in Table~\ref{VP}. Each video has different resolutions; thus, they vary in size, which will affect processing and transmission time. Since the Raspberry has limited computational and memory capacity, the frame size will impact the performance significantly.

\begin{table}[h]
\renewcommand{\arraystretch}{1.5}
\caption{Video Properties}
\label{VP}
\centering 
\begin{tabular}{|c|c|c|}
\hline
 version & resolution & size  \\
\hline
\hline
 v1 & 256x134 & 156 KB  \\
\hline
\hline
 v2 & 426x224 & 350 KB  \\
\hline
\hline
 v3 & 640x338 & 576 KB  \\
\hline
\hline
 v4 & 854x450 & 1.31 MB  \\
\hline
\hline
 v5 & 1280x674 & 2.73 MB  \\
\hline
\hline
 v6 & 1920x1012 & 5.71 MB  \\
\hline
\end{tabular}
\end{table}

\begin{figure*}
\begin{centering}
\begin{tabular}{ccc}
\includegraphics[keepaspectratio=true,angle=0,width=62mm]{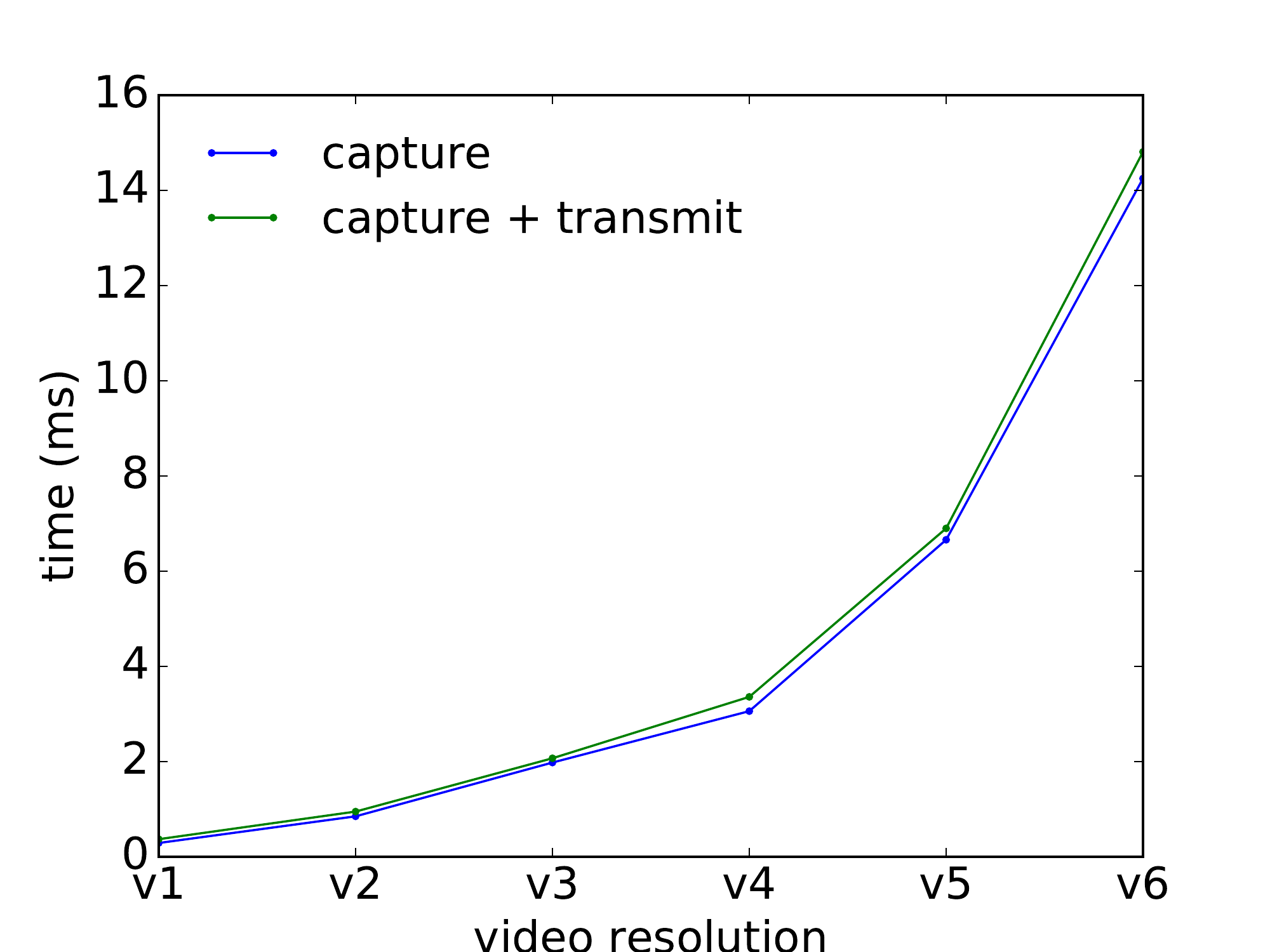} &
\hspace{-9mm}
\includegraphics[keepaspectratio=true,angle=0,width=62mm]{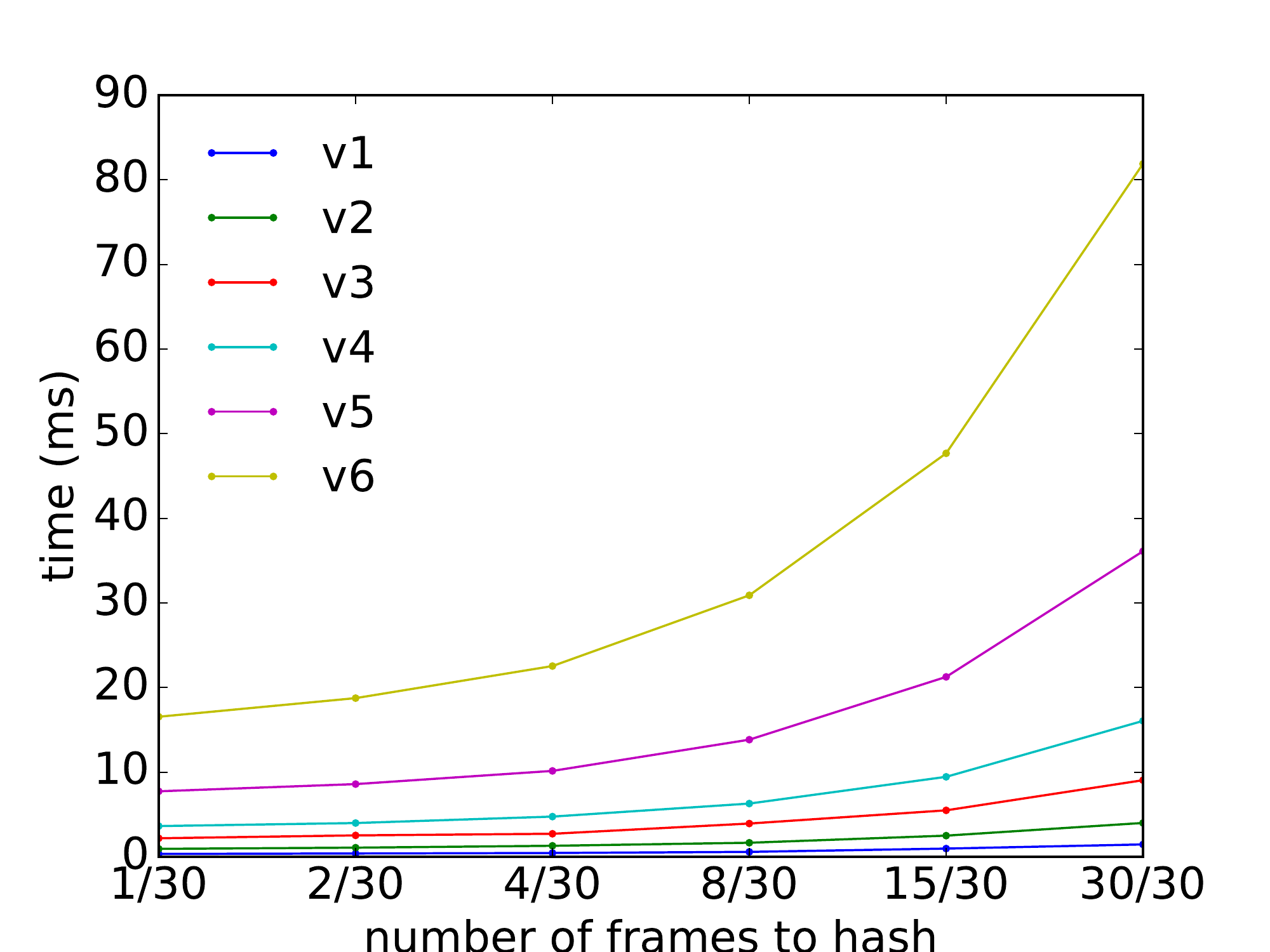} &
\hspace{-9mm}
\includegraphics[keepaspectratio=true,angle=0,width=62mm]{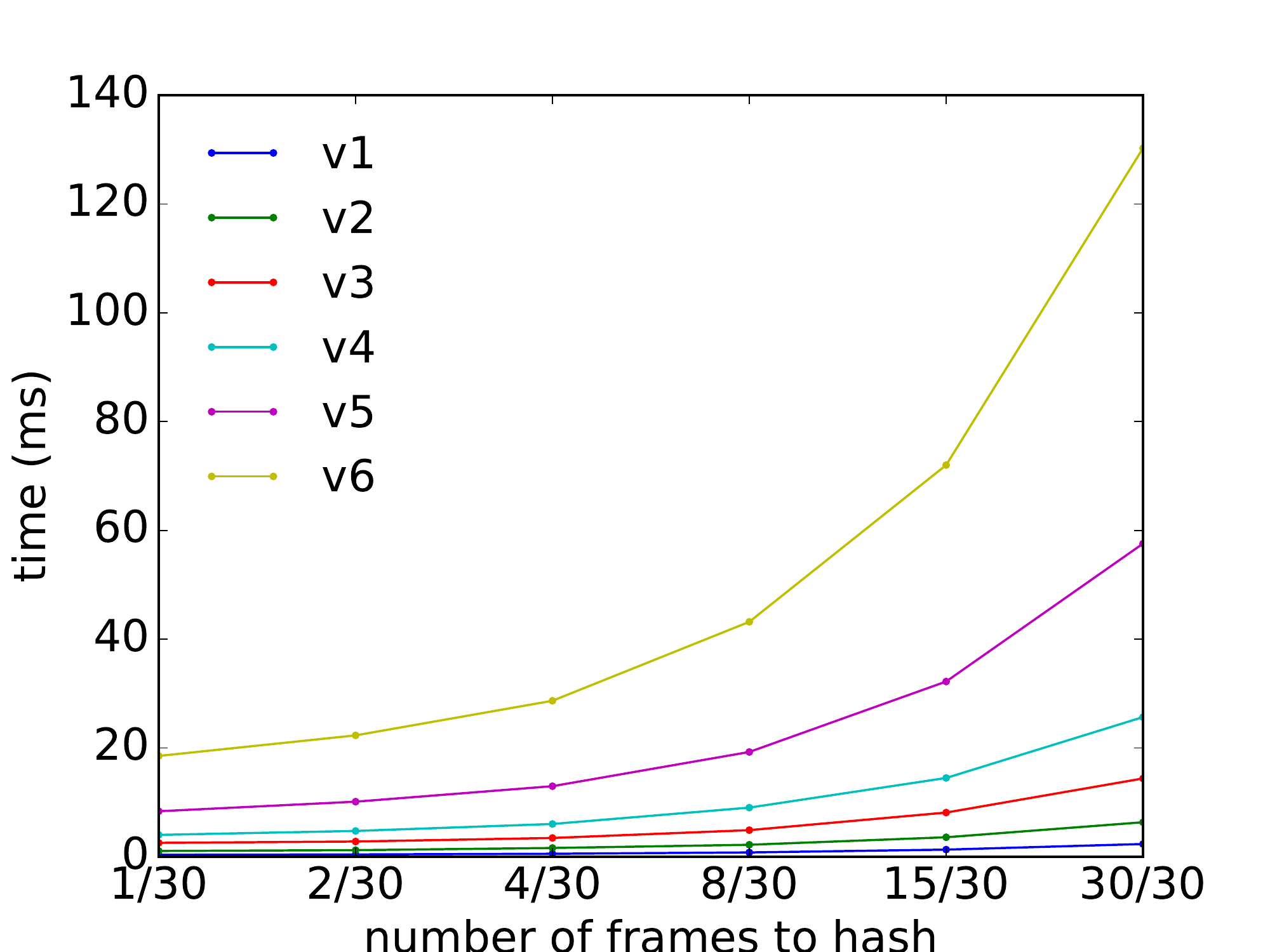}  

 \tabularnewline
\footnotesize{ a. Frame Capture and Transmission} &
\hspace{-9mm}
\footnotesize{ b. String Conversion } & 
\hspace{-9mm}
\footnotesize{ c. Hashing Process }

 \tabularnewline

\includegraphics[keepaspectratio=true,angle=0,width=62mm]{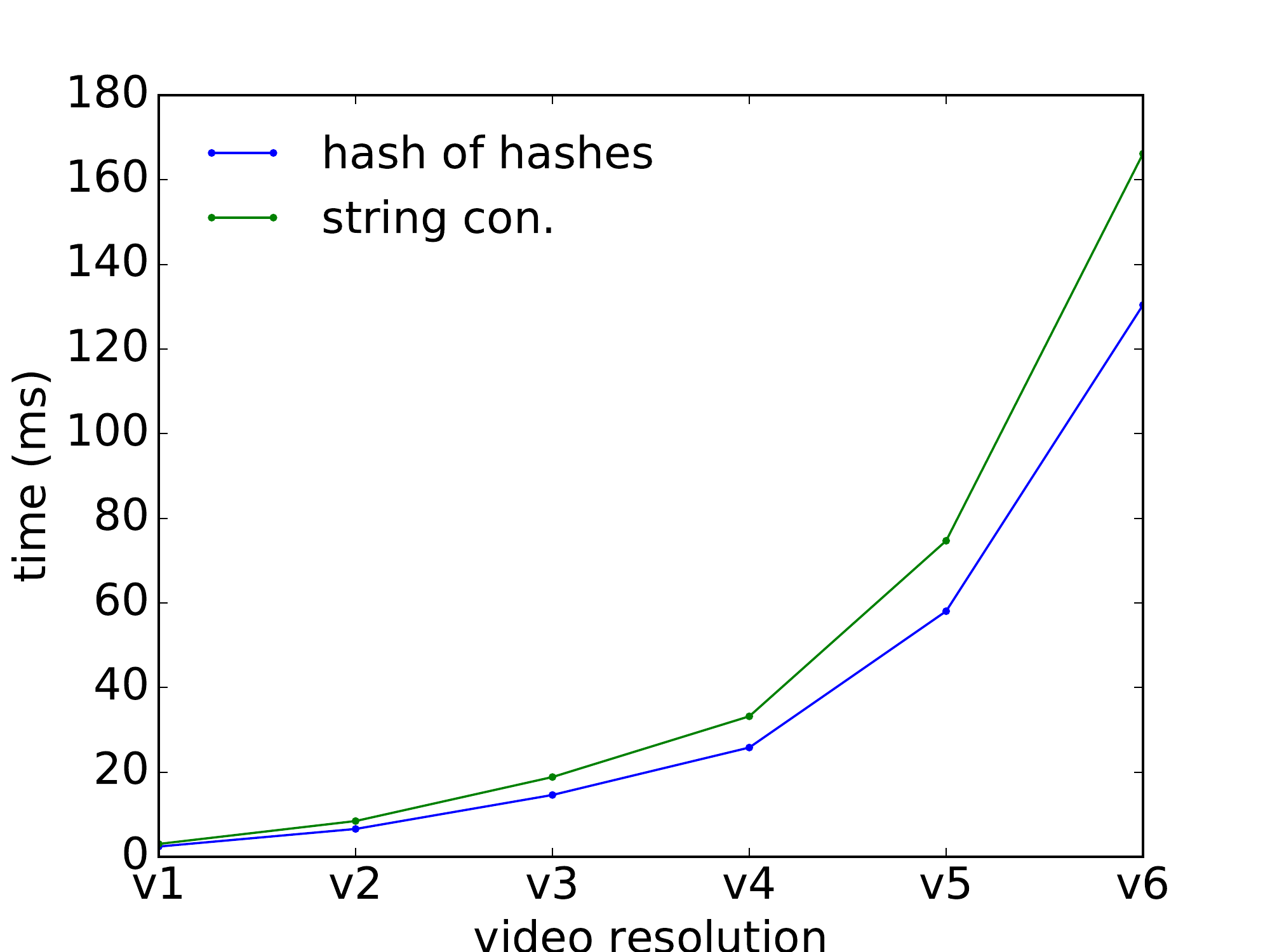} & 
\hspace{-9mm}
\includegraphics[keepaspectratio=true,angle=0,width=62mm]{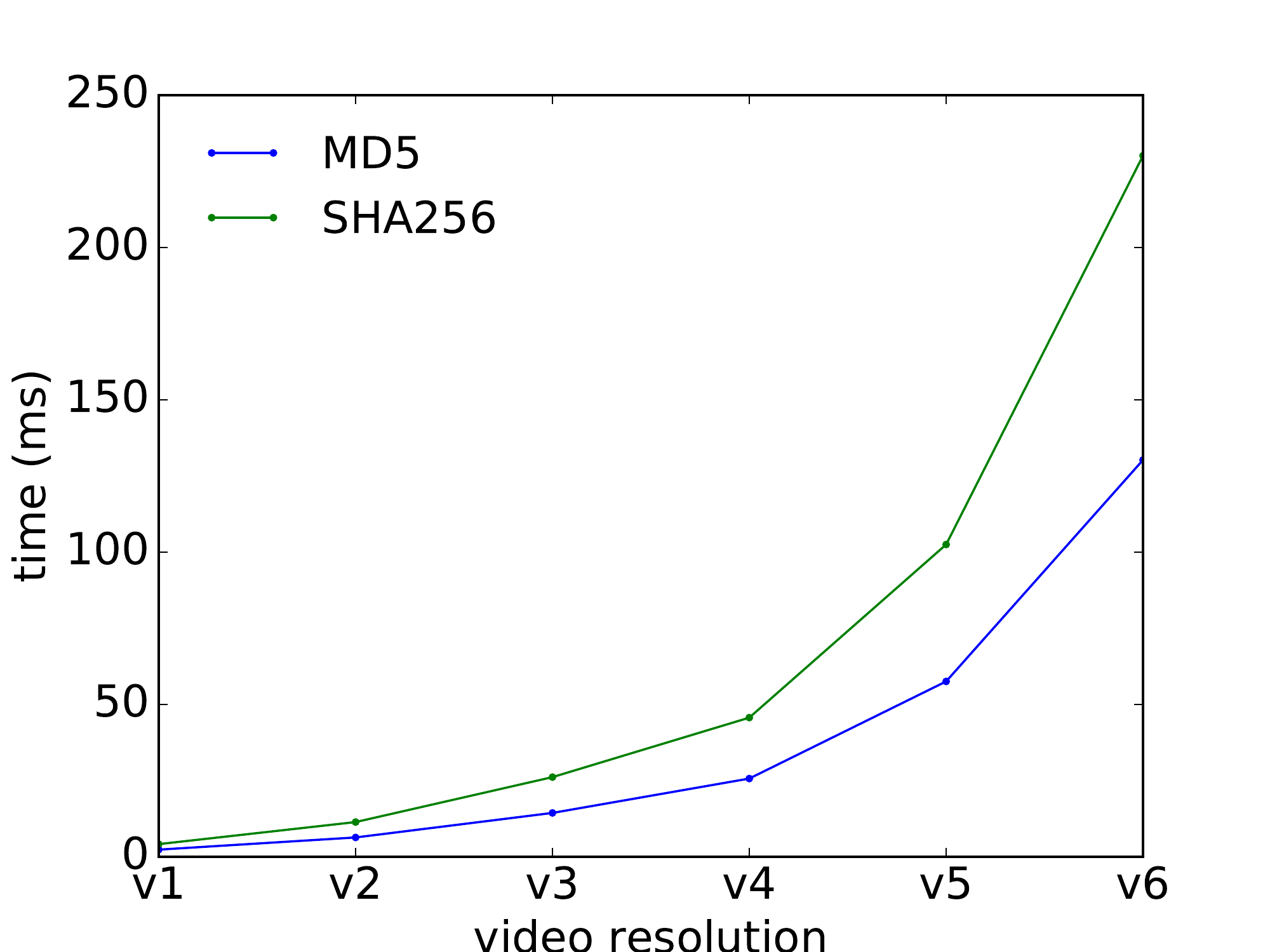} & 
\hspace{-9mm}
\includegraphics[keepaspectratio=true,angle=0,width=62mm]{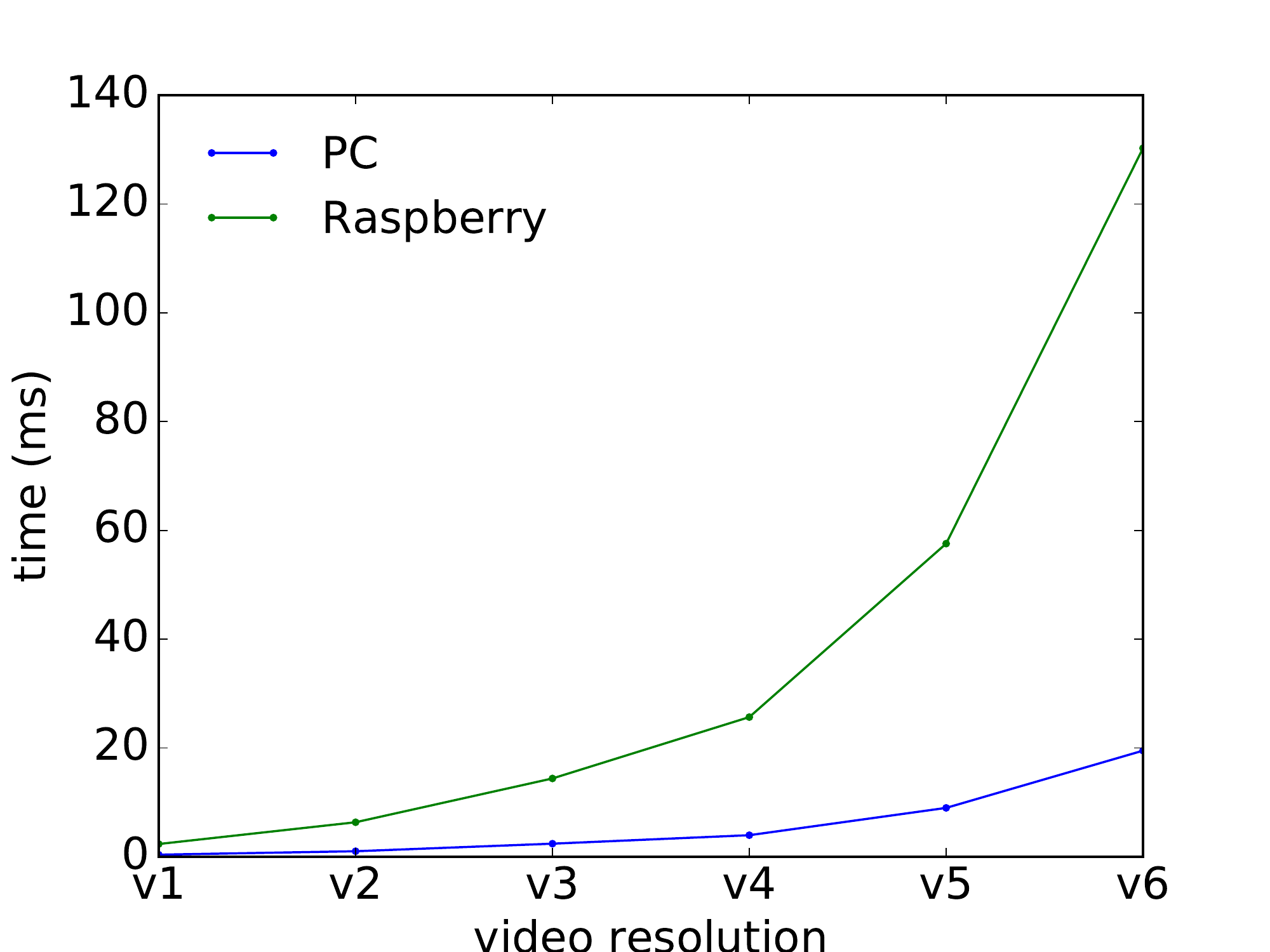}  

 \tabularnewline
\footnotesize{ d. Alternative Blcokchain Writing Methods} & 
\hspace{-10mm}
\footnotesize{ e. Different Hash Function } &
\hspace{-10mm}
\footnotesize{ f. Performance on PC } 
 \tabularnewline
\end{tabular}
\caption{Experiment Results }
\label{fig:fig2}
\vspace{-5mm}
\end{centering}
\end{figure*}

\subsection{Metrics and Baselines}

The main metric we are concerned with is the time to process and stream the videos. We considered both the computation time on the IoT device and communication time. The other metric is the resolution of the videos used. 

As a baseline, we considered different numbers of frames to be hashed. Specifically, frequency of frames ranging from 1/30 (one frame from each second) to 30/30(each frame) are considered. We also compared our approach on Raspberry PI to that on a more powerful machine such as a desktop.

\subsection{Performance Results}

First, we assessed the video capturing and transmission time of the Raspberry without having any additional processes running on it. Fig~\ref{fig:fig2}.a shows the time needed for a Raspberry to capture and transmit the frames to a server over WiFi. It took 14.25 seconds to capture a 10 second video in the highest quality which means that there will be pauses when we stream it. This indicates that it is not possible to record video in this resolution using a Raspberry even with no other processes slowing it down. A lower resolution video will help the hardware handle the process smoothly. Transmitting the frame from IoT device to a remote station is the next step. 
The transmission time depends on the quality of the communication channel. Current congestion level, protocol, distance, transmission environment are some factors that can effect the time. In our case, the IoT device connects to the server through a router using WiFi. We tried UDP and TCP connections and did not change other parameters, as that is not the purpose of this study. Both UDP and TCP gave very similar results as shown in Fig.~\ref{fig:fig2}.a and this is almost negligible compared to computation overhead for capturing the video. It should be noted that the video is in mp4 format. Another encoding technique might generate a different size for the same quality which will effect the transmission time.

The next thing we analyzed is string conversion and the hashing process. 
Fig~\ref{fig:fig2}.b shows the time for capturing and converting the frames to string for each video. It seems conversion to string is an intensive process which adds a lot to the total time, especially for larger frames. If all frames are processed, the total time for the highest resolution reaches around 80 seconds. Only the lowest 3 resolutions are under 10 seconds. If we consider processing only selected frames instead of each frame, then the task can be completed faster. If only two frames each second are converted, then all the resolutions except the highest one can be handled. The hashing process involves first converting the frame to a string, and then taking the hash of the string.  

Table~\ref{TM} lists approximate times for conversion and hashing of a single frame based on the frame size. It takes almost half a second to convert and hash a video frame in 1920x1012. The results indicate that hashing takes less time once string conversion is done. Fig~\ref{fig:fig2}.c shows the total time needed for all the tasks together. Low resolution videos (v1=256x134 and v2=426x224) are handled within the time limit even when all the frames are processed, but higher resolutions require frame selection which lets us go all the way up to v5 without exceeding the time limit. If all frames are needed to check integrity, we can achieve it only for for v1 and v2.

\begin{table}[!t]
\renewcommand{\arraystretch}{1.5}
\caption{Processing Times}
\label{TM}
\centering
\begin{tabular}{|c|c|c|}
\hline
 resolution & conversion(ms) & hashing(ms)  \\
\hline
\hline
 256x134 & 3.8 & 2.9 \\
\hline
\hline
 426x224 & 10.3  & 7.7 \\
\hline
\hline
 640x338 & 23.3  & 17.6 \\
\hline
\hline
 854x450 & 42.9  & 31.7 \\
\hline
\hline
 1280x674 & 97.1  & 70.8 \\
\hline
\hline
 1920x1012 & 223.2  & 159.6 \\
\hline
\end{tabular}
\end{table}
  
Recall that there was another concern to to take into consideration. Storing hash of every frame in hyperledger might be costly and unattractive even though we are able to do so. Hence, we tested two other options: 1) appending strings from each frame so that we will have only one string to be hashed for every 30 frames; 2) appending the hash of each frame to be hashed and stored in blockchain. That means, we connect to the blockchain once for every 30 frames. Basically here, we take all the frames into consideration in two different ways, but we only need to write to blockchain occasionally. We tested these two options to see the performance, as is shown in Fig~\ref{fig:fig2}.d. We can achieve both options for v1 and v2 under 10 seconds while the second option performs a little better because the string does not get too big. 

We also considered using different hash functions. The experiments up to now have been performed using MD5. We wanted to test another one to see if it makes a difference. Thus, we employed SHA256 instead of MD5. The results Fig~\ref{fig:fig2}.e shows the total time using two different hashing functions while processing each frame. MD5 outperforms SHA256 in terms of speed. The difference indicates that developing lightweight specific hashing algorithms for video enables higher quality videos to be transmitted.

The last thing we wanted to investigate is how the whole process would run on a regular computer, which has more powerful resources to improve performance. We ran the program on a machine with i5 quadcore CPU and 6 GB RAM.  Fig~\ref{fig:fig2}.f shows the comparison for the highest resolution, v6, with every frame processed. The performance on PC, especially for higher resolutions, is much better than on Raspberry. If the idea is applied on higher capacity machines for any surveillance purposes, it can allow for higher quality recording.

\section{Conclusion}
\label{sec:conclusion}

In this paper, we proposed and implemented a system to verify the integrity of a video captured by wireless IoT devices by incorporating blockchain. The idea was for a video that is being streamed, to generate the hash values for individual frames and write those hashes in a distributed ledge, before the frames are transmitted. 
We implemented the system in a real setup with Hyperledger as the blockchain technology. The experiment results indicate that the idea is promising and usable as long as the right video resolution is picked. Further investigation is needed to handle higher resolution videos.

\section*{Acknowledgment}
Dominik Danko in this work is supported by US National Science Foundation under the grant number CNS-REU-1757761. This work is also supported in part by \#NPRP9-257-1-056 grant from the Qatar National Research Fund. 

\bibliographystyle{IEEEtran}

\end{document}